\documentclass{elsarticle}

\usepackage{amsmath,multirow}

\biboptions{compress}

\newcommand{\ed}{ 
\bibliography{ABC_citation,LiteraturSM_2016,QCD-5-1,dfg}
\end{document}
}

\newcommand{\ice}[1]{\relax}

\newcommand{\re}[1]{(\ref{#1})}

\newcommand{\ba}{\begin{eqnarray}}
\newcommand{\ea}{\end{eqnarray}}
\newcommand{\beq}{\begin{equation}}
\newcommand{\eeq}{\end{equation}}

\ice{

\newcommand{\beq}{\begin{equation}}
\newcommand{\eeq}{\end{equation}}

\newcommand{\EQN}{\label}
\newcommand{\ovl}{\overline}
}

\usepackage[latin1]{inputenc} 
\usepackage[T1]{fontenc}
\usepackage[safe]{textcomp}
\usepackage{lmodern} 
\usepackage{epsfig}
\usepackage{float}
\usepackage{amstext}  
\usepackage{amsfonts}
\usepackage{amssymb} 
\usepackage{amsmath}
\usepackage{amsthm}
\usepackage{bm}       
\usepackage[bottom]{footmisc} 
\usepackage{array}                
\usepackage{xcolor} 
\usepackage{framed}
\usepackage{slashed}
\usepackage[absolute]{textpos}
\usepackage{axodraw4j}
\usepackage{dsfont}
\usepackage{multirow}

\usepackage{appendix}
\usepackage{listings}
\usepackage{enumerate}     
\usepackage[bottom]{footmisc} 
\usepackage{array}           
\usepackage{parskip}        
\usepackage{xcolor} 
\usepackage{color}
\usepackage{framed}
\usepackage{fancyhdr}
\usepackage{subfig}

\usepackage{tikz}
\usepackage[absolute]{textpos}
\usepackage{pstricks}
\usepackage{axodraw4j}
\usepackage{graphicx}

\newcommand{\p}{\partial}

\newcommand{\f}[2]{\frac{#1}{#2}}
\newcommand{\sss}[1]{\scriptscriptstyle{#1}}
\newcommand{\ssst}[1]{\scriptscriptstyle{\text{#1}}}

\newcommand{\bea}{\begin{eqnarray}}
\newcommand{\eea}{\end{eqnarray}}
\newcommand{\be}{\begin{equation}}
\newcommand{\ee}{\end{equation}}
\newcommand{\beas}{\begin{eqnarray*}}
\newcommand{\eeas}{\end{eqnarray*}}
\newcommand{\bes}{\begin{equation*}}
\newcommand{\ees}{\end{equation*}}
\newcommand{\bas}{\begin{align*}}
\newcommand{\eas}{\end{align*}}

\newcommand{\ssL}{{\mathcal L}}

\newcommand{\cf}{C_{\scriptscriptstyle{F}}} 
\newcommand{\ca}{C_{\scriptscriptstyle{A}}}

\newcommand{\gs}{g_{\scriptscriptstyle{s}}}

\newcommand{\lb}{\left(}
\newcommand{\rb}{\right)}

\newcommand{\msbar}{$\overline{\text{MS}}$}

\ice{
 перенес (временно?) в основной текст
\newcommand{\dFFfiNRex}{\frac{d_{\scriptscriptstyle{F,r}}^{abcd}d_{\scriptscriptstyle{F,i}}^{abcd}}{\dFr}}
\newcommand{\dFFfiNR}{\tilde{d}^{\sss{(4)}}_{\sss{FF,ri}}}
\newcommand{\dFAfNRex}{\frac{d_{\scriptscriptstyle{F,r}}^{abcd}d_{\scriptscriptstyle{A}}^{abcd}}{\dFr}}
\newcommand{\dFAfNR}{\tilde{d}^{\sss{(4)}}_{\sss{FA,r}}}
}

\newcommand{\cfi}{C_{\scriptscriptstyle{F,i}}}

\newcommand{\cfr}{C_{\scriptscriptstyle{F,r}}} 

\newcommand{\tri}{T_{\scriptscriptstyle{F,i}}}

\newcommand{\trr}{T_{\scriptscriptstyle{F,r}}}

\newcommand{\Nfi}{n_{\scriptscriptstyle{f,i}}}

\newcommand{\Nfr}{n_{\scriptscriptstyle{f,r}}}

\definecolor{bluemar}{rgb}{0,0,.5}
\definecolor{redmar}{rgb}{.8,0,0}
\definecolor{greenmar}{rgb}{0,.5,0}

\newcommand{\bigrint}{%
\hbox to 2.92em{\hss\scalebox{1.1}[1] {\rotatebox[origin=c]{15}{$\displaystyle\int$}}\hss}}

\catcode`\@=11
\def\slash{\mathpalette\make@slash}
\def\make@slash#1#2{\setbox\z@\hbox{$#1#2$}%
  \hbox to 0pt{\hss$#1/$\hss\kern-\wd0}\box0}
\catcode`\@=12 % @ signs are no longer letters

\def\nnb{\nonumber}

\newcommand{\sbz}{\,}
\newcommand{\EQN}[1]{\hspace{3mm}\fbox{\fbox{$#1$}} \label{#1}}

\renewcommand{\EQN}[1]{\mlabel{#1}}

\renewcommand{\EQN}[1]{\label{#1}}

\newcommand{\al}{\alpha}
\newcommand{\ovl}[1]{\overline{#1}}

\newcommand{\dd}{{\rm d}}
\def\1{\hbox{{1}\kern-.25em\hbox{l}}}% Web of Conferences font
\ice{
\setlength{\textwidth}{17cm}
\setlength{\oddsidemargin}{1.cm}

\setlength{\headsep}{1.0cm}
\setlength{\topmargin}{0.0cm}
\setlength{\topskip}{0.1cm}
\setlength{\footskip}{1.5cm}
\frenchspacing
}

\begin{document}

\begin{frontmatter}
  \title{
Adler function, Bjorken Sum Rule and  \\ Crewther-Broadhurst-Kataev relation
 with generic fermion representations at  order $O(\alpha^4_s)$
\makebox(0,0){\raisebox{5cm}{\hspace{5.5cm}  TTP22-039    }}
}
\author{K.\,G.~Chetyrkin}
\ead{konstantin.chetyrkin@partner.kit.edu}
\address{Institut f\"ur Theoretische Teilchenphysik,
Karlsruher Institut f\"ur Technologie, Karlsruhe, Germany}
\begin{abstract}
 We compute the nonsinglet Adler $D$-function and the coefficient function
 for Bjorken polarized sum rules $S^{Bjp}$ at order $O(\alpha_s^4)$ in an
 extended QCD model with arbitrary number of fermion representations. 
 The Crewther-Broadhurst-Kataev (CBK) relation
% \cite{Crewther:1972kn}, 
% \cite{Broadhurst:1993ru}
 in this order is confirmed.
\ice{
 As a byproduct we also get the Larin's
 correction factor $Z^f_{ns}$ for the nonsiglet axial curent at the same order
 as well as the corresponding anomalous dimension at order $O(\alpha_s^5)$.
}
\end{abstract}

\end{frontmatter}

\section{Introduction}
\label{intro}

The Crewther-Broadhurst-Kataev (CBK) relation
\cite{Crewther:1972kn,Broadhurst:1993ru}
demonstrates  a non-trivial connection between two
(at first sight  seemingly unrelated)
important  physical quantities,  namely the 
(non-singlet) Adler  $D$-function
\beq
D(L,a) = 1 
+ 3\,C_F\, a +
\sum_{i=2}^{\infty} \  {d_i(L)} \, a^i(\mu^2)
\EQN{D}
\eeq
and  the (non-singlet)  coefficient function for  the  Bjorken polarized sum
rules 
\beq
S^{Bjp}(L,a) =1  - 3\,C_F\, a +
\sum_{i=2}^{\infty} \  {c_i(L)}\, a^i(\mu^2)  
\EQN{bjp1}
{}.
\eeq
Here $L=\ln \frac{\mu^2}{Q^2}$, $\mu$ is the normalization scale in \msbar-scheme
\cite{tHooft:1973mm,Bardeen:1978yd}
(which we will assume  throughout the  paper) and $a=\frac{g_s^2}{16 \pi^2}=\frac{\alpha_s}{4\pi}$
(precise definitions for both  functions  and color  factors involved  will  be given in
Sections  \ref{adler}, \ref{bjorken} and \ref{prel1}  correspondingly).

The functions \re{D} and \re{bjp1} are  very   well studied in perturbative QCD. Due
to works
\cite{Chetyrkin:1979bj,Dine:1979qh,Celmaster:1979xr,Gorishnii:1990vf,Chetyrkin:1996ez,Baikov:2008jh,%
Herzog:2017dtz,Gorishnii:1983gs,Larin:1990zw,Baikov:2010je}
they are  known to impressively high order $\alpha_s^4$.  The CBK relation connecting both
functions reads\footnote{We  omit direct indication on $L$-dependence in  places where   it  can not
lead  to  misunderstandings.} :
\beq 
{D}(a)\,{C}^{Bjp}(a)
= 1 + \beta(a)\, K(a) ,  
\ \ \
  K(a) = 
a\,K_1 +a^2\,K_2 +a^3\,K_3 
+ \dots 
\label{Crewther} 
\eeq
Here
\beq
\beta(a) = \mu^2\,\frac{\mathrm{d}}{\mathrm{d} \mu^2}
\ln a(\mu) =\sum_{i \ge 1} \beta_i a^{i}
\eeq
is the QCD
$\beta$-function describing the {\em running} of the coupling constant
$a$ with respect to a  change of the normalization scale $\mu$ and
with its first term 
\[
\beta_1 = -\frac{11}{3}\, C_A + \frac{4}{3} T_F \,n_f
\]
being responsible for {asymptotic freedom} of QCD.
The
term proportional the $\beta$-function responsible for  deviation from
the limit of exact conformal invariance, with the deviation starting in order $\alpha_s^2$,  
and was
suggested \cite{Broadhurst:1993ru} on the basis of 
${\cal O}(\alpha_s^3)$ calculations of $D(a)$ \cite{Gorishnii:1990vf,Surguladze:1990tg} 
and ${C}^{Bjp}(a)$  \cite{Larin:1991tj}.
The original  relation without this term was first proposed  in
\cite{Crewther:1972kn}.

The  fact that the CBK relation is valid up to maximally  known order in $\alpha_s$
is highly non-trivial. Indeed, a  simple counting of  available  color factors
shows that fulfillment of  \re{Crewther} sets as many as 6 constraints at the sum
$d_4 + c_4$ and all of them are met identically. At lower orders the number of constraints 
is 2 and 3 for the sums $d_2 + c_2$ and  $d_3 + c_3$  correspondingly (see discussions
in \cite{Broadhurst:1993ru,Baikov:2010je} and Section \ref{cbk}).

Some formal arguments in favour of \re{Crewther} were suggested in
\cite{Crewther:1997ux,Braun:2003rp}. Unfortunately, these considerations can
not replace a real proof. Such a proof should demonstrate at least how it
works in detail and in which renormalization schemes it holds\footnote{It has
been shown in \cite{Garkusha:2011xb} that the CBK relation ceases to take
place in the {}'t Hooft \msbar-based scheme.}. Finally, it would be highly
desirable if the future proof would clarify a way of computing the factor
$K(a)$ {\em directly} that is without previous calculations of $D(a)$
and ${C}^{Bjp}(a)$.

In the  present work we use an extended QCD (eQCD) model with arbitrary number of
fermion representations in order to subject the CBK relation to one more
non-trivial test. We compute both components $D(a)$ and ${C}^{Bjp}(a)$
within the extended QCD to order $\alpha_s^4$ and demonstrate the validity of
the resulting CBK relation. Let us stress that the knowledge of both $D(a)$
and ${C}^{Bjp}(a)$ in QCD with multiple  fermion representations provides
important ingredients to obtain the so-called $\beta$-expansion representation
\cite{Mikhailov:2004iq,Mikhailov:2016feh,Kataev:2010du,Kataev:2014jba} for
observables.  This representation allows one to apply the extended BLM (eBLM)
approach to optimize the PT series
\cite{Mikhailov:2004iq,Kataev:2014jba,Kotlorz:2018bxp}.  The approach suggests
a way to resum the non-conformal parts of various QCD observables into the
scale of the coupling in a unique way for any optimization task.  Note that
there exists an alternative  method known as the Principle of Maximum
Conformality (PMC)
\cite{Brodsky:2011ta,Brodsky:2011ig,Brodsky:2013vpa}.  In this approach
content of $\beta$-expansion as well as  results of optimization in general
differ from those  in eBLM.

\ice{for the use of so-called extended BLM approach
  \cite{Mikhailov:2004iq,Mikhailov:2016feh,Kataev:2010du,Kataev:2014jba}.  The
  approach suggests a way for resumming the non-conformal parts of various QCD
  observables into the scale of the coupling in a unique way\footnote{
    Note  that there exists an alternative approach known as the Principle of
    Maximum Conformality (PMC)
    \cite{Brodsky:2011ta,Brodsky:2011ig,Brodsky:2013vpa}.
    }.}

%It should be noted that the above mentioned proofs
%are suggestive considerations

%are two very well studied 

\section{Preliminaries}
\label{prel}

\subsection{QCDe Lagrangian and notations for  color factors}
\label{prel1}
The Lagrangian of a (massless) QCD-like model extended to include several
fermion representations of the gauge group (to be  referred as  QCDe) is given by
(our notations essentially follow  those of \cite{Zoller:2016sgq})
\bea
\ssL_{\sss{QCD}}&=&-\f{1}{4}G^a_{\mu \nu} G^{a\,\mu \nu}-\f{1}{2 \lambda}\lb\p_\mu A^{a\,\mu}\rb^2 
+\p_\mu \bar{c}^a \p^{\mu}c^a+\gs f^{abc}\,\p_\mu \bar{c}^a A^{b\,\mu} c^c \nonumber \\
&+&\sum\limits_{r=1}^{N_{\ssst{rep}}}\sum\limits_{q=1}^{\Nfr}
\left\{\f{i}{2}\bar{\psi}_{q,r}\overleftrightarrow{\slashed{\p}}\psi_{q,r}
%-m_{q,r}\bar{\psi}_{q,r}\psi_{q,r}
+ \gs \bar{\psi}_{q,r}\slashed{A}^a T^{a,r} \psi_{q,r}\right\}{},
\label{LQCD} 
\eea  
with \ice{the gluon field strength tensor and }
\be
G^a_{\mu \nu}=\p_\mu A^a_\nu - \p_\nu A^a_\mu + \gs f^{abc}A^b_\mu A^c_\nu{},
\  \   \
\left[ T^{a,r},T^{b,r} \right]=if^{abc}T^{c,r}
{},
\ee
and $f^{abc}$ being the structure constants of the gauge group.  The index $r$
specifies the fermion representation and the index $q$ the fermion flavour,
$\psi_{q,r}$ is the corresponding fermion field\ice{and $m_{q,r}$ the
corresponding fermion mass}. The number of fermion flavours in
representation $r$ is $\Nfr$ for any of the $N_{\ssst{rep}}$ fermion
representations.

For every fermion  representation $r$  we have two  quadratic Casimir operators
$\cfr$ and $\trr$
  \be   \delta_{ij} \cfr = T^{a,r}_{ik} T^{a,r}_{kj},
\ \ \
\trr \delta^{ab}=\textbf{Tr}\lb T^{a,r} T^{b,r}\rb=T^{a,r}_{ij} T^{b,r}_{ji}
{}.
\ee
The dimension of $r$  will be denoted as $d_{F,r}$. As for gluon (adjoint) representation we use
the standard notation $C_A$ and $N_A$ for the corresponding  quadratic Casimir operator and
dimension of the gluon representation. The  standard QCD  corresponds to  the case of
$N_{\ssst{rep}} = 1$. If $N_{\ssst{rep}} > 1$ we will consider  the first fermion representation as
a  special  one in what follows  with
\[
C_{F,1} \equiv C_F,\  \  d_{F,1} \equiv  d_F, \ \  n_{f,1} \equiv n_f, \ \
T_{F,1} \equiv T_F \  \
\mbox{and} \ \  T^{a,1} \equiv T^a
{}.
\]
Let us stress that all external  operators (like the EM current) which appear later are assumed to
involve only fermion fields $\psi_{q,1}$ which we will refer also as $\psi_{q}$.  

In  addition to quadratic Casimir operators we need also  quartic ones which are expressed in terms of
symmetric  tensors (see \cite{COLOR} for  details)
\be d_{\sss{R}}^{a_1 a_2 a_3  a_4}=\f{1}{n!} 
\sum\limits_{\text{perm } \pi}
\text{Tr}\left\{ T^{a_{\pi(1)},R}T^{a_{\pi(2)},R} T^{a_{\pi(3)},R}   T^{a_{\pi(4)},R}\right\}{}, \label{dRa1an}
\ee
where $R$ can be any fermion representation, $R= {F,r} $ ($ r=1 \dots  N_{\ssst{rep}} $)    or the adjoint representation, $R=A$  where
$T^{a,A}_{bc}=-i\,f^{abc}$.

\newcommand{\dFAfNR}{\tilde{d}_{\sss{FA}}}

\newcommand{\dFAfNRex}{\frac{d_{\scriptscriptstyle{F}}^{abcd}d_{\scriptscriptstyle{A}}^{abcd}}{d_{\sss{F}}}}

\newcommand{\dFFfiNR}{\tilde{d}_{\sss{FF,r}}}
\newcommand{\dFFfiNRex}{\frac{d_{\scriptscriptstyle{F}}^{abcd}d_{\scriptscriptstyle{F,r}}^{abcd}}{ d_{\sss{F}}  }}

The following quartic Casimir operators appear in our results at order $\alpha_s^4$:
\be
\dFAfNR =  \dFAfNRex,\quad  \dFFfiNR=\dFFfiNRex,
\EQN{quartic}
\ee
with $d_{\scriptscriptstyle{F}}^{abcd} \equiv d_{\scriptscriptstyle{F,1}}^{abcd}$  and
$\tilde{d}_{FF} \equiv \frac{d_{\scriptscriptstyle{F}}^{abcd}d_{\scriptscriptstyle{F,1}}^{abcd}}{ d_{\sss{F}}  }$.
%It is useful to  keep in mind that in QCD instead of 

 \subsection{Adler function in QCDe}
\label{adler}

We start with  the (non-singlet) polarization function $\Pi(L,a)$ of the 
vector current $j_{\al} = \ovl{\psi}_q\gamma_{\al}\psi_q$ and defined as
\newcommand{\NS}{\mathrm{NS}}
\newcommand{\MS}{\mathrm{MS}}
\beq
 (-g_{\al\beta} q^2 + q_{\al} q_{\beta}) \, \Pi^{\NS}(L,a) =
i\int\dd^4 x \, e^{iq\cdot x}\langle 0|{\rm  T}j_\al(x)j_\beta(0)|0\rangle^{\NS}
{},  
\label{Pi}
\eeq
where    $Q^2 = -q^2$,
$L = \ln \frac{\mu^2}{Q^2}$. It is understood that the rhs of \re{Pi} includes
only non-singlet diagrams, that is those  with  both external currents belonging to
one and the same quark loop.
The Adler function is defined as  (the normalization factor below
is conventionally fixed by the requirement that in Born approximation
the Adler function starts from one)
\cite{Adler:1974gd} 
\beq
d_F\, n_f\, {D}(L,a_s) =  -12\, \pi^2\,
Q^2\, \frac{\mathrm{d}}{\mathrm{d} Q^2} \Pi^{\NS}(L,a)
{}.
\label{D:def}
\eeq It is worthwhile to note that the Adler function unlike the polarization
operator \re{Pi} is scale invariant due to the derivative in $Q^2$ which kills
a quadratic UV divergence of \re{Pi} around integration region $x \approx 0$.

\subsection{Bjorken  function in QCDe}
\label{bjorken}

The most convenient for us  definition of the coefficient function $S^{Bjp}$
comes from the following  Operator Product Expansion (OPE):
\ba
&&
\int T{[ j_{\alpha}(x)\, j_{\beta}(0)]}\, e^{iqx}\, dx|_{q^2\rightarrow{-\infty}}
\approx  
\label{OPE}
\frac{q^{\sigma}}{q^2} 
C^{Bjp}_{\NS}(L,a)\,
 \epsilon_{\alpha\beta\rho\sigma} \,
 A_{\rho}(0)
 \\
 &&
 +\dots \ \mbox{(singlet and  other terms)}
\nnb
 %+\dots
{},
\ea
where 
$
A_{\rho} = \ovl{\psi}_q \, \gamma_{\rho} \gamma_5  \,\psi_q
%\EQN{axial}
$
is the axial vector  current. The function $C^{Bjp}_{\NS}$ is by definition contributed
by non-singlet diagrams only
%only while all  other terms (singlet and higher twist ones) are
(see Fig. 1). In what follows we will not  write index $\NS$ explicitly.
Singlet contributions to OPE \re{OPE} were discussed in  \cite{Larin:2013yba}.
\begin{center}
\begin{figure}
\hspace{15mm}\includegraphics[width=33mm, angle=-90]{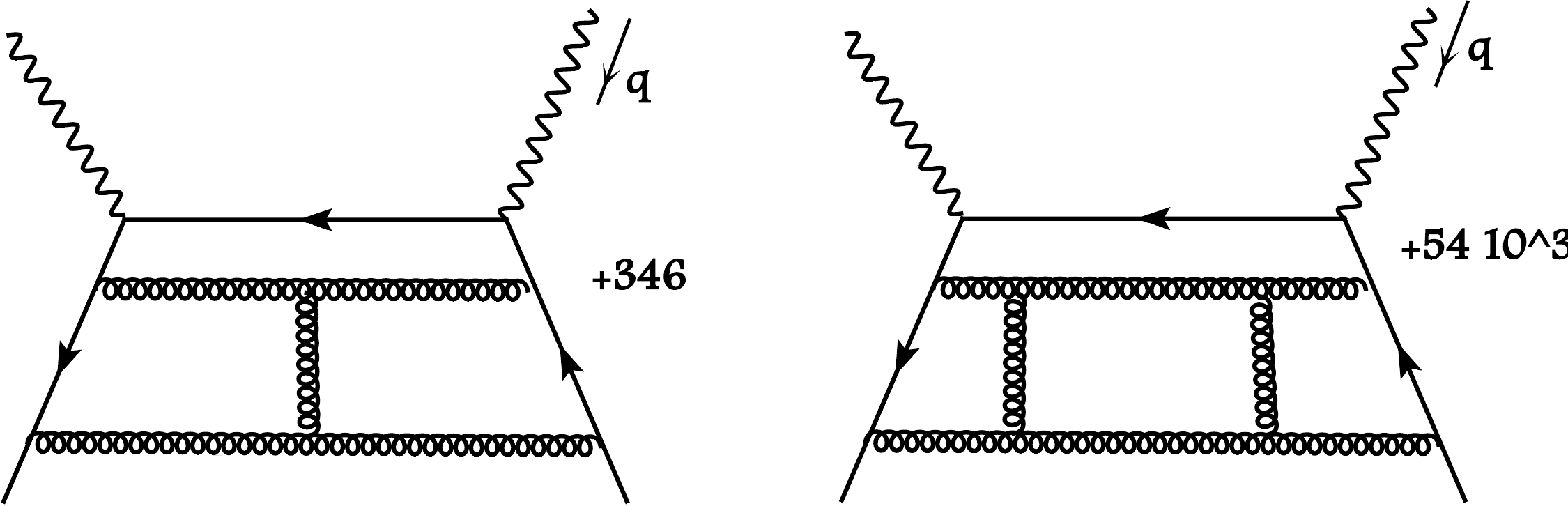}
\caption{Examples of diagrams contributing to  the coefficient function  $C^{Bjp}_{NS}$   at three
and four loops. }
\label{CBjp:NS}
\end{figure}
\end{center}
There is a technical subtlety in definition of the $\gamma_5$ matrix appearing in the
definition of the axial current.  Following works  \cite{Larin:1991tj,Baikov:2010je}  we will use so-called Larin's approach
\cite{Larin:1993tq}.
It means that renormalized axial current is defined as
\beq
A_{\rho}  \equiv  z_A^{\NS} \, \frac{1}{6}\epsilon_{\alpha\beta\sigma\rho} \, 
[\, \ovl{\psi}_q \gamma_{[\alpha\beta\sigma]} \psi_q\,]^{\MS},
\ \ \gamma_{[\alpha\beta\sigma]} \equiv \frac{1}{2}
%\left(
(
\gamma_{\al}
\gamma_{\beta}
\gamma_{\sigma}
-
\gamma_{\sigma}
\gamma_{\beta}
\gamma_{\al}
)
%right)
\EQN{larin}
{},
\eeq
where 
$[\, \ovl{\psi}_q \gamma_{[\alpha\beta\sigma]} \psi_q\,]^{\MS}$ stands for the
\msbar-renormalized current. The (finite) factor   $z_A^{\NS}$ is  chosen
in such a way to effectively restore the anticommutativity of $\gamma_5$
(see corresponding discussion in \cite{Larin:1993tq,Chetyrkin:1993hk}). 

\subsection{Color factors}
%{  in QCD  and  QCDe}

For future reference let us describe color factors which appear in all three
components of the CBK relation\ice{ as a result of explicit
  calculations}. First, we note an obvious fact that one and the same
collection of color factors may appear in $d_n$, $c_n$.  Second, due to the
prefactor $\beta(a)$ in \re{Crewther}, the same set of color factors describes
coefficient $K_{n+1}$.  This is true for both QCD and QCDe cases
\cite{Baikov:2010je}. Another important fact is that transition from QCD to QCDe does  touch
only $n_f$-dependent color  factors\footnote{This means that contributions
proportional to $n_f$-independent color factors are identical in both
cases.}. 
The corresponding modifications are shown
in Table 1. Here we use the following  notations:
\ice{
\ba \label{eq:d.o.f.}
\textbf{nT}&=&\sum\limits_{i} \Nfi \tri,\quad
\textbf{nTCk}= \sum\limits_{i} \Nfi \tri  \cfi^k,\quad \textbf{n}\bm{d^{abcd}}= \sum\limits_{i}\Nfi d^{abcd}_{F,i}\,.
\ea
}
\beq
\textbf{nT} \equiv \sum\limits_{i} \Nfi \tri,\quad \textbf{nTC1}  \equiv\sum\limits_{i} \Nfi \tri  \cfi,
\quad \textbf{nTC2} \equiv \sum\limits_{i} \Nfi \tri  \cfi^2
{}.
\eeq
Note that  if a color structure in  the left column of  Table 1
does not proliferate then the   corresponding contributions  should be identical
in QCD and QCDe results.

In  order to transform a QCDe result to  the corresponding one
in the standard QCD one should
make the following replacements:
\ba
&\textbf{nT} \to n_f T_F,\quad \textbf{nTC1}
\to  n_f T_F C_F,
\EQN{qcde.2.qcd.rules.1}
\nnb
\\
& \quad \textbf{nTC2} \to   n_f T_F C_F^2,
\,
\displaystyle{
\sum_r \, n_{\sss{f,r}}  \dFFfiNR  \to n_f \tilde{d}_{\sss{FF}}
}
             {}.
\EQN{qcde.2.qcd.rules.2}             
\ea
\renewcommand{\arraystretch}{1.4}
\begin{center}
\begin{table}[t] % put at top of page if possible 
\begin{center}
  \begin{tabular}{c|c}
    \hline
QCD & QCDe \\
\hline
\multicolumn{2}{c}{$\alpha_s^2$} 
\\
\hline
%\\
$n_f T_f$ &  $\textbf{nT}$  \\
\hline
%\\
\multicolumn{2}{c}{$\alpha_s^3$} 
\\
\hline
%\\
$C_F^2 n_f T_f$   & $ C_F^2 \textbf{nT}, \ C_f\,   \textbf{nTC1}$
\\
$C_F n_f^2 T_f^2$   & $ C_F  (\textbf{nT})^2$
\\
$C_F C_A\, n_f T_f$   & $ C_F  C_A \, \textbf{nT}$
\\
\hline
%\\
\multicolumn{2}{c}{$\alpha_s^4$} 
\\
\hline
$C_F^3 n_f T_f$   & $ C_F^3 \textbf{nT}, \ C_f^2\,   \textbf{nTC1},  \ C_f\,   \textbf{nTC2} $
\\
$C_F^2 n_f^2 T_f^2$   & $ C_F  (\textbf{nT})^2,  C_f\,   \textbf{nTC1} $
\\
$C_F^2 C_A\, n_f T_f$   & $ C_F^2  C_A \, \textbf{nT},  C_F C_A \, \textbf{nTC1} $
\\
$C_F n_f^3 T_f^3$    & $ C_F  (\textbf{nT})^3 $
\\
$C_F C_A  n_f^2 T_f^2$    & $ C_F C_A \, (\textbf{nT})^2 $
\\
$C_F C_A^2  n_f T_f $    & $ C_F C_A^2  (\textbf{nT}) $
\\
 $ n_f \tilde{d}_{\sss{FF}}$ &  $\displaystyle \sum_r \, n_{\sss{f,r}} \, \dFFfiNR $
\\
\hline
\end{tabular}
\caption{Proliferation of the $n_f$-dependent color factors in the QCDe-model.  \label{table} }
\end{center}
\end{table}
\end{center}

An inspection of Table 1 clearly shows that in the case of the QCDe-model the
number of extra constraints imposed by the CBK relation on the
combinations $d_3 + c_3$ and $d_4 +c_4$ is increased from 3 and 6 to 4 and 9
correspondingly.

\ice{
As the results in QCDe  have to coincide
with

The following

%It is instructive and  useful  to compare

Let us consider general structure of color factors which may appear in all
three components of the CBK relation. The following statements
}

\section{Calculation and results}
\label{calc}
\subsection{Results for $D$ and  $S^{Bjp}$ }

We have computed the functions $D$ and $S^{Bjp}$ to order ${\cal{O}}(\alpha_s^4)$
using essentially the same methods as in \cite{Baikov:2010je} (for a short
review see \cite{Baikov:2015tea}). All momentum diagrams have beem generated
with QGRAF \cite{QGRAF} and reduced to master integrals (well known from
\cite{Baikov:2010hf,Lee:2011jt}) with the help of the $1/D$ expansion \cite{Baikov:2005nv,Baikov:1996rk}.
  
For calculation of color factors we have employed a generalization of the FORM
\cite{Vermaseren:2000nd:old} package COLOR \cite{COLOR} developed by M.~Zoller
\cite{Zoller:2016sgq}.  Below we present our results for the 
Adler fuction and the coefficient function $S^{Bjp}$ as defined by
(\ref{D},\ref{OPE}). Note that we set \mbox{$\mu^2=Q^2$}; the full dependence
on $\mu$ can be easily restored  by expressing $a(Q^2)$ by $  a(\mu^2)$ with
the help of the standard RG evolution equation for $a$ (the $\beta$-function
for QCDe is known at four loops from \cite{Zoller:2016sgq}).
\ba
d_1&=& 3\cf,
\EQN{d1} \\
d_2&=&-\frac{3}{2}\cf^2+ \cf \ca \left(\frac{123}{2}-44 \zeta_3\right)-2\cf (\textbf{nT})(11-8 \zeta_3),
\EQN{d2}
\\
d_3&=&-\frac{69}{2} \cf^3
\nnb
+
\\
&&
\cf^2\bigg[\ca \left(-127-572  \zeta_3+880 \zeta_5\right)+ (\textbf{nT})(72+ 208
    \zeta_3-320  \zeta_5)\bigg] + \nonumber\\
&&\cf \ca^2 \left(\frac{90445}{54}-\frac{10948}{9}\zeta_3-\frac{440}{3}\zeta_5\right)+
\nnb
\\
&&
\cf\ca (\textbf{nT}) \left(-\frac{31040}{27}+\frac{7168}{9} \zeta_3+\frac{160}{3} \zeta_5\right)+\nonumber\\
    &&\cf(\textbf{nT})^2
   \left(\frac{4832}{27}-\frac{1216}{9}\zeta_3\right)+\cf(\textbf{nTC1}) (-101+96 \zeta_3), 
\EQN{d3}
\ea
\ba
d_4&=&\cf^4 \left(\frac{4157}{8}+96 \zeta_3\right)+ \nonumber \\
   && \cf^3 \Big[\ca (-2024-278 \zeta_3+18040
  \zeta_5-18480 \zeta_7)
\nnb
  \\
  &&
  \hspace{12mm}-
  \textbf{nT}\, (-298 +56 \zeta_3+6560 \zeta_5-6720\zeta_7)\Big]+ \nonumber \\
   &&\cf^2 \bigg[\ca^2\! \left(\!-\frac{592141}{72}\!-\frac{87850}{3}\zeta_3+\frac{104080}{3}\zeta_5+9240
   \zeta_7\right)\!+ \nonumber \\
   &&\phantom{\cf^2 \bigg[}\ca (\textbf{nT})\! \left(\frac{67925}{9}+\frac{61912}{3}\zeta_3-\frac{83680}{3}\zeta_5-3360 \zeta_7
     \right)+\nonumber
     \ea
     \ba
%//     
     &&\phantom{\cf^2 \bigg[}(\textbf{nT})^2
   \left(-\frac{13466}{9}-\frac{10240}{3} \zeta_3+\frac{16000}{3} \zeta_5\right)+
   \textbf{nTC1}
   (251+ 576 \zeta_3-960 \zeta_5)\bigg]+\nonumber \\
   && \cf \Bigg[\ca^3 \left(\frac{52207039}{972}-\frac{912446}{27} \zeta_3-\frac{155990}{9} \zeta_5+4840
   \zeta_3^2-1540
   \zeta_7\right) \nonumber \\
   &&\phantom{\cf \bigg[} \ca^2 (\textbf{nT}) \left(-\frac{4379861}{81}+\frac{275488}{9}\zeta_3+\frac{150440}{9}\zeta_5-1408
   \zeta_3^2+560 \zeta_7\right)+  \nonumber \\
   &&\phantom{\cf \bigg[}  \ca (\textbf{nT})^2
     \left(\frac{1363372}{81}-\frac{83624}{9} \zeta_3-\frac{43520}{9}\zeta_5-128 \zeta_3^2\right)
     +  \nonumber
     \\
     & &\phantom{\cf \bigg[ } \ca (\textbf{nTC1}) \left(
      - \frac{375193}{54}
      + 7792 \zeta_3
      +400 \zeta_5
      -2112 \zeta_3^2        \right)\!
       +\nonumber
       \\
   &&\phantom{\cf \bigg[}\!(\textbf{nT})^3 \left(-\frac{392384}{243}+ 
   \frac{25984}{27}\zeta_3 + \frac{1280}{3} \zeta_5\right)+ \nonumber
%   \ea
%   \ba
\\
   & &\phantom{\cf \bigg[ }(\textbf{nT}) (\textbf{nTC1})
  \left(\frac{63250}{27}-2784 \zeta_3+768 \zeta_3^2\right)+
\nnb
  \\
  &&
  \hspace{7mm}
  \textbf{nTC2} \left(\frac{355}{3}+
   272 \zeta_3-480 \zeta_5\right)\Bigg]-%\right)
   \nonumber \\
   && 16\left[\sum\limits_{r} n_{f,r} \,\tilde{d}_{\text{FF},r}\cdot\left(
     13
     +
     16\zeta_3
     -
    40\zeta_5\right)+\tilde{d}_\text{FA}\cdot\left(-3+ 4\zeta_3+20\zeta_5\right)\right]\,.
\EQN{d4}
   \ea
The results for $c_k$ of the Bjorken SR in QCDe,
\ba
c_1&=& -3\cf,
\EQN{c1}
\\
c_2&=&\frac{21}{2} \cf^2 - 23 \ca \cf + 8 \cf (\textbf{nT}),
\ea
 \ba
c_3&=&-\frac{3}{2} \cf^3+ 
\cf^2 \left[\ca
  \left(\frac{1241}{9}-\frac{176}{3} \zeta_3\right)-\textbf{nT} \left(\frac{664}{9}-\frac{64}{3}\zeta_3\right)\right]+
\nnb
\\
&&
\cf\ca^2 \left(-\frac{10874}{27}+\frac{440}{3}\zeta_5\right)+
\nonumber \\
 &&
\cf\ca(\textbf{nT}) \left(\frac{7070}{27}+48 \zeta_3-\frac{160}{3} \zeta_5\right)
-
\nnb
\\
&&
\cf (\textbf{nT})^2\frac{920}{27}+ \cf (\textbf{nTC1}) (59-48 \zeta_3),                                                     \ea
\ba
c_4&=&-\cf^4 \left(\frac{4823}8 + 96 \zeta_3\right)  +   \\
    &&\phantom{-}\cf^3 \left[ -
    \ca \left(\frac{3707}{18} + \frac{7768}3 \zeta_3 - \frac{16720}3 \zeta_5\right)\right.+ 
\nnb
    \\
    &&
    \hspace{10mm}
    \left.\textbf{nT} \left(\frac{5912}{9} + \frac{3296}3 \zeta_3 - \frac{6080}3 \zeta_5\right)\right]+ \nonumber \\
 && \phantom{-}\cf^2 \bigg[
    \ca^2 \left(\frac{1071641}{216} + \frac{25456}{9} \zeta_3 - \frac{22000}{9} \zeta_5 - 6160 \zeta_7\right) -\nonumber \\
 &&\phantom{ -\cf^2 \bigg[ }   \ca (\textbf{nT}) \left(\frac{106081}{27} + \frac{9104}{9} \zeta_3 - \frac{8000}{9} \zeta_5 - 2240 \zeta_7\right)+ \nonumber \\
 &&\phantom{ -\cf^2 \bigg[ } (\textbf{nT})^2 \left(\frac{16114}{27} - \frac{512}{3} \zeta_3\right) - \textbf{nTC1}
    \left(\frac{1399}{3} - 400 \zeta_3\right)\bigg] + \nonumber \\
 && \cf \Bigg[ \ca^3 \left(-\frac{8004277}{972} + \frac{4276}{9} \zeta_3 + \frac{25090}{9} \zeta_5 
    - \frac{968}{3} \zeta_3^2 +
       1540 \zeta_7\right) + \nonumber \\
&&\phantom{\cf \Bigg[}   \ca^2 (\textbf{nT}) \left(\frac{1238827}{162} + 236 \zeta_3 - \frac{14840}{9} \zeta_5 + \frac{704}{3} \zeta_3^2-
       560 \zeta_7\right) - \nonumber \\       
 &&\phantom{\cf \Bigg\{}  
       \ca(\textbf{nT})^2 \left(\frac{165283}{81} + \frac{688}{9} \zeta_3 - \frac{320}{3} \zeta_5+ \frac{128}{3} \zeta_3^2\right)+\nonumber \\
       &&\phantom{\cf \Bigg\{} \ca (\textbf{nTC1}) \left(\frac{124759}{54} - 1280 \zeta_3 - 400 \zeta_5\right)+\nonumber
       \\
       &&
\nnb \phantom{\cf \Bigg\{}
       \frac{38720}{243} (\textbf{nT})^3 - (\textbf{nT})(\textbf{nTC1}) \left(\frac{19294}{27} - 480 \zeta_3\right)
       -
       \\
       && \phantom{\cf \Bigg\{}
       \textbf{nTC2} \left(\frac{292}3 + 296 \zeta_3 - 480 \zeta_5\right)      
       \Bigg] + \nonumber \\
  & & 16\left[
         \sum\limits_{r} n_{f,r} \,\tilde{d}_{\text{FF},r}\cdot\left(
         13
         +
         16\zeta_3
         -
         40\zeta_5\right)+\tilde{d}_\text{FA}\cdot\left(-3+ 4\zeta_3+20\zeta_5\right)
   \right].
       \nnb
\ea
\ice{
\subsection{Results for  Larin's factor and anomalous dimension of
 $[\, \ovl{\psi}_q \gamma_{[\alpha\beta\sigma]} \psi_q\,]^{\MS}$
}
}
\section{CBK relation in QCDe  }
\label{cbk}

Using the color structures of $d_2$ and  $d_3$  as  templates we  find that the CBK relation \re{Crewther}
is indeed fulfilled identically with the following values  for the coefficients $K_i$:

\ba
K_1&=&
\cf\left(  -\frac{21}{2}+12 \sbz \zeta_{3}  \right)
, \\
K_2&=&
\cf^2 \left( \frac{397}{6}+136 \sbz \zeta_{3}-240 \sbz \zeta_{5}  \right)
+ \cf \ca
\left(  -\frac{629}{2}+\frac{884}{3} \sbz \zeta_{3}  \right)
\nnb
\\
&& +
(\cf\textbf{nT})
\left( \frac{326}{3}-\frac{304}{3} \sbz \zeta_{3}  \right)
{},
\ea
\ba
K_3&=&
\cf^3
\left( \frac{2471}{12}+488 \sbz \zeta_{3}-5720 \sbz \zeta_{5}+5040 \sbz \zeta_{7}  \right)
\nnb
\\
&&
+\cf^2\ca
\left( \frac{99757}{36}+\frac{16570}{3} \sbz \zeta_{3}-\frac{24880}{3} \sbz \zeta_{5}-840 \sbz \zeta_{7}  \right)
 \nonumber\\
 &&
+\cf \ca^2
\left(
-\frac{406043}{36} + \frac{72028}{9} \sbz \zeta_3 - 1232 \sbz \zeta_3^2 + \frac{11900}{3} \sbz \zeta_5
\right)
\nnb
\\
&&
+\cf\ca (\textbf{nT})
\left( \frac{67520}{9}-\frac{40336}{9} \sbz \zeta_{3}-\frac{8000}{3} \sbz \zeta_{5}-128 \,\zeta_3^2  \right)
\nonumber\\
&&
+
\cf(\textbf{nT})^2
\left(  -\frac{9824}{9}+\frac{6496}{9} \sbz \zeta_{3}+320 \sbz \zeta_{5}  \right)
\nnb
\\
&& \hspace{0mm}
+ \cf^2(\textbf{nT})
\left(  -\frac{11573}{9}-2288 \sbz \zeta_{3}+4000 \sbz \zeta_{5}  \right)
\nnb
\\
&&
+
\cf(\textbf{nTC1})  \left( \frac{1713}{2}-1380 \sbz \zeta_{3}+576 \,\zeta_3^2  \right)
{}.
\ea

As expected from Table 1 and relations \re{qcde.2.qcd.rules.1} and
\re{qcde.2.qcd.rules.2} coefficient $K_2$ in QCD is essentially identical to
the one in QCD (that is after identification $\textbf{nT}$ with $n_f
T_f$). Coefficient $K_3$ in QCDe is different from the case of QCD only by 2
last terms. All constraints imposed by the CBK relation are fulfilled.

\section{Conclusion} 
\label{concl}

We have computed the nonsinglet Adler $D$-function and the coefficient function
for Bjorken polarized sum rules $S^{Bjp}$ at order $O(\alpha^4)$ in the
extended QCD model.  The CBK  relation is confirmed.

These  results  have been extensively used for construction and analyzing
explicit expressions for the elements of the $\{ \beta \}$-expansion for
the nonsinglet Adler $D$-function and Bjorken polarized sum rules $S^{Bjp}$
in the  N$^4$LO and higher orders in \cite{beta:exp:22}.

They may be also useful for renormalization group analysis of
the $D$ and  $S^{Bjp}$  functions  in large-$N_c$ and large-$N_f$ limits 
\cite{Ryttov:2018uue,Girmohanta:2019cth}.

For readers's convenience  all our results are collected  in an ancillary file. 

\section*{Acknowledgments}

Author is grateful to P.\,A.~ Baikov and S.\,V.~Mikhailov for essential help,
useful comments and good advice.  I would like to express my special thanks to
M.~Zoller for his kind permission to use his extension of the package COLOR.

%and  numerous explanations of its  features.
%patient help in my work  with it.

The work  was supported in part by DFG grant CH~1479/2-1.

%{ABC:22,beta:exp:22,Braun:2022byg,Chetyrkin:2021qvd,Braun:2021cqe,Braun:2020ymy,Baikov:2019zmy}

\end{document}